\documentclass[12pt,thmsa]{article}
\usepackage{amsfonts}

\input{tcilatex}
\begin{document}

\author{Harald Grosse$^a$, Karl-Georg Schlesinger$^b$ \qquad  \\
$^a$Institute for Theoretical Physics\\
University of Vienna\\
Boltzmanngasse 5\\
A-1090 Vienna, Austria\\
e-mail: grosse@doppler.thp.univie.ac.at\\
$^b$Institute for Theoretical Physics\\
University of Vienna\\
Boltzmanngasse 5\\
A-1090 Vienna, Austria\\
e-mail: kgschles@esi.ac.at}
\title{The universal envelope of the topological closed string BRST-complex}
\date{}
\maketitle

\begin{abstract}
We construct a universal envelope for any Poisson- and Gerstenhaber algebra.
While the deformation theory of Poisson algebras seems to be partially
trivial, results from string and $M$-theory suggest a rich deformation
theory of Gerstenhaber algebras. We apply our construction in this case to
well known questions on the topological closed string BRST-complex. Finally,
we find a simliar algebraic structure, as for the universal envelope, in the 
$SU\left( 2\right) $-WZW model.
\end{abstract}

\section{Introduction}

The BRST-complex of the topological closed string is known to have the
structure of a differential Gerstenhaber algebra (see \cite{HM} and the
literature cited therein). In the first part (section 2) of this paper, we
show how to construct a universal envelope for any Poisson and Gerstenhaber
algebra. The result is a trialgebra in the sense of \cite{CF} and a graded
version thereof (which we will call an odd super-trialgebra for reasons to
be discussed below), respectively. This immediately implies the existence of
a universal envelope for the topological closed string BRST-complex.
Moreover it follows that the deformation theory of the odd-super trialgebra
corresponds to the third bicomplex (the Gerstenhaber complex) of \cite{HM}
if the coproduct of the odd super-trialgebra is kept fixed. Using recent
results of \cite{LMN}, we derive an argument from this suggesting that the
conjecture of \cite{HM} on the physical interpretation of the Gerstenhaber
complex as relating to $\left( 1,1\right) $ little string theory on the NS5
brane in type IIB string theory should reduce in a dual description on the $%
N=2$ supersymmetric sine-Gordon model coupled to topological gravity to a
categorified version of the Kazhdan-Lusztig theorem on representation theory
of affine Lie algebras.

In section 3, we discuss several examples of Poisson algebras. We show that
in all these cases the deformation theory of the Poisson algebra reduces to
that of the underlying ordinary associative algebra or is trivial. One is
always forced to consider the bracket not as independent algebraic structure
but as the first order deformation of the product. This is the usual
interpretation of Poisson brackets in deformation theory, of course, but for
the abstract notion of Poisson algebra this gives a severe restriction on
the deformation theory. We conjecture that such a property should generally
hold true for Poisson algebras for mathematics, as well, as physics
motivated reasons. This implies that the deformation theory of trialgebras,
arising as universal envelopes of Poisson algebras, is heavily restricted.
In contrast, the deformation theory of Gerstenhaber algebras implies that
odd super-trialgebras have a rich deformation theory, i.e. superextensions
seem to be a decisive ingredient for trialgebraic structures. This suggests
that four dimensional topological field theories might only lead to
nontrivial invariants of 4-manifolds if they possess \textit{in this sense}
a kind of supersymmetry (We would like to stress that we only refer to these
special graded algebraic structures, here. Obviously, it does not make sense
to speculate of supersymmetry in the usual sense, referring e.g. to a
Lagrangian formulation, in this context, since topological field theories
do, in general, not even satisfy the spin-statistics theorem).

In section 4, we consider strings moving on the $S^3$ in the transversal
geometry of a flat NS5 brane which are described by the $SU\left( 2\right) $%
-WZW model. We find that also in the $SU\left( 2\right) $-WZW model an odd
super-trialgebra appears, defined by the bicovariant differential calculus
on the $q$-deformed fuzzy sphere.

Section 5 contains some concluding remarks.

\bigskip

\section{The universal envelope of Poisson- and Gerstenhaber algebras}

Recall that a Poisson algebra is defined as a commutative, associative
algebra $\left( A,\cdot \right) $ equipped with a Lie bracket $\left[
,\right] $ such that for all $a,b,c\in A$ 
\[
\left[ a,b\cdot c\right] =\left[ a,b\right] \cdot c+b\cdot \left[ a,c\right] 
\]
Similarly, a Gerstenhaber algebra is defined as a $\Bbb{Z}_2$-graded
associative algebra $\left( A,\cdot \right) $ with graded commutative even
product $\cdot $ and an odd Lie bracket $\left[ ,\right] $ such that for all 
$a,b,c\in A$ 
\[
\left[ a,b\cdot c\right] =\left[ a,b\right] \cdot c+\left( -1\right)
^{\left( |a|-1\right) |b|}b\cdot \left[ a,c\right] 
\]

\bigskip

\begin{remark}
Observe that in contrast to the case of super-Lie algebras or graded Poisson
algebra, the Lie bracket of a Gerstenhaber algebra is odd with respect to
the grading and, correspondingly, the derivation property is odd from the
sign rule.
\end{remark}

\bigskip

\begin{definition}
A trialgebra $(A,*,\Delta ,\cdot )$ with $*$ and $\cdot $ associative
products on a vector space $A$ and $\Delta $ a coassociative coproduct on $A$
is given if both $(A,*,\Delta )$ and $(A,\cdot ,\Delta )$ are bialgebras and
the following compatibility condition between the products is satisfied for
arbitrary elements $a,b,c,d\in A$: 
\[
(a*b)\cdot (c*d)=(a\cdot c)*(b\cdot d)
\]
\end{definition}

\bigskip

Trialgebras were first suggested in \cite{CF} as an algebraic means for the
construction of four dimensional topological quantum field theories. It was
observed there that the representation categories of trialgebras have the
structure of so called Hopf algebra categories (see \cite{CF}) and it was
later shown explicitly in \cite{CKS} that from the data of a Hopf category
one can, indeed, construct a four dimensional topological quantum field
theory. The first explicit examples of trialgebras were constructed in \cite
{GS 2000a} and \cite{GS 2000b} by applying deformation theory, once again,
to the function algebra on the Manin plane and some of the classical
examples of quantum algebras and function algebras on quantum groups. In 
\cite{GS 2001} it was shown that one of the trialgebras constructed in this
way appears as a symmetry of a two dimensional spin system. Besides this,
the same trialgebra can also be found as a symmetry of a certain system of
infinitely many coupled $q$-deformed harmonic oscillators.

With the notion of a trialgebra at hand, we can now formulate a concept of a
universal envelope of a Poisson algebra:

Let $\left( A,\cdot ,,\left[ ,\right] \right) $ be a Poisson algebra. Let $%
\widehat{U}\left( A\right) $ be the universal envelope of the Lie algebra $%
\left( A,\left[ ,\right] \right) $ completed with respect to the inclusion
of formal power series. We can obviously extend the product $\cdot $ from $A$
to the tensor algebra over $A$ (and to a completion of the tensor algebra by
formal power series) by 
\begin{equation}
\left( a\otimes b\right) \cdot \left( c\otimes d\right) =\left( a\cdot
c\right) \otimes \left( b\cdot d\right)  \label{1}
\end{equation}
for all $a,b,c,d\in A$.

\bigskip

\begin{lemma}
The product defined by (\ref{1}) induces a product on $\widehat{U}\left(
A\right) $ and together with the commutative Hopf algebra structure on $%
\widehat{U}\left( A\right) $ this product gives the structure of a
trialgebra to $\widehat{U}_G\left( A\right) $ where we define $\widehat{U}%
_G\left( A\right) $ to be given as the subalgebra generated by the
group-like elements in $\widehat{U}\left( A\right) $.
\end{lemma}

\proof%
Recall that a group-like element in $\widehat{U}\left( A\right) $ is defined
by the property 
\[
\Delta \left( a\right) =a\otimes a
\]
The relations of a trialgebra are checked by calculation. Essential is the
fact that the quotient relations of $\widehat{U}\left( A\right) $ do involve
three elements of $A$ at once which means that they do not interfere with
the compatibility relation (\ref{1}) of a trialgebra.

Finally, one has to check that the derivation property of a Poisson algebra
does not interfere with the structure of a trialgebra. For this, observe
that upon identifying 
\begin{equation}
c\leftrightarrow c\cdot 1+1\cdot c  \label{2}
\end{equation}
and 
\[
c\otimes 1\leftrightarrow 1\otimes c\leftrightarrow c 
\]
for all $c\in A$, the derivation property can be derived from the
compatibility relations of the trialgebra. Namely, 
\begin{eqnarray*}
\left[ a\cdot b,c\right] &=&\left[ a\cdot b,c\cdot 1+1\cdot c\right] \\
&=&\left( a\cdot b\right) \otimes \left( c\cdot 1+1\cdot c\right) -\left(
c\cdot 1+1\cdot c\right) \otimes \left( a\cdot b\right) \\
&=&\left( a\otimes c\right) \cdot \left( b\otimes 1\right) -\left( c\otimes
a\right) \cdot \left( 1\otimes b\right) \\
&&+\left( a\otimes 1\right) \cdot \left( b\otimes c\right) -\left( 1\otimes
a\right) \cdot \left( c\otimes b\right) \\
&=&\left[ a,c\right] \cdot b+a\cdot \left[ b,c\right]
\end{eqnarray*}
where $\otimes $ denotes the product in $\widehat{U}\left( A\right) $.
Observe that the above identification does not interfere with the fact that
nontrivial trialgebras are always non-unital (see \cite{GS 2000b}) since we
have not used a normalization in (\ref{2}).

This concludes the proof. 
\endproof%

\bigskip

In a completely similar way, we get a super-trialgebra $\widehat{U}_G\left(
A\right) $ from a graded Poisson algebra $\left( A,\cdot ,\left[ ,\right]
\right) $ and an odd super-trialgebra from a Gerstenhaber algebra where
super-trialgebras and odd super-trialgebras have the obvious gradings. By
construction of the universal envelope, an odd super-trialgebra has one of
the two products of odd type with respect to the grading and the
compatibility relation of the products obeys an odd sign rule. In
conclusion, we have the following result:

\bigskip

\begin{lemma}
The universal envelope $\widehat{U}_G\left( A\right) $ of a Gerstenhaber
algebra $\left( A,\cdot ,\left[ ,\right] \right) $ has the structure of an
odd super-trialgebra.
\end{lemma}

\bigskip

Here, as we will do in the sequel, we have called $\widehat{U}_G\left(
A\right) $ the universal envelope of the Gerstenhaber (or Poissson-) algebra.

\bigskip

The BRST-complex of the topological closed string has the structure of a
differential Gerstenhaber algebra, i.e. a Gerstenhaber algebra with a
differential $d$ obeying 
\[
d^2=0 
\]
(strictly speaking, only the cohomology has the structure of a differential
Gerstenhaber algebra while the complex is only a homotopy Gerstenhaber
algebra but in a first approach we will neglect BRST exact terms in this
paper which means that we can treat the comlex as a differential
Gerstenhaber algebra, too). Lemma 2 immediately implies that the universal
envelope of the BRST-complex of the topological closed string has the
structure of an odd super-trialgebra with differential.

In \cite{HM} it was shown that the deformation theory of the BRST-complex of
the topological closed string is described by three different bicomplexes:
First, there is the complex of $A_\infty $-deformations ($d$ and $\cdot $
are deformed but $\left[ ,\right] $ is kept fixed) which is generated by
deforming the closed string correlation functions by closed string bulk
operators. This is the complex described by the WDVV-equations. Next, one
has the complex of $L_\infty $-deformations ($d$ and $\left[ ,\right] $ are
deformed but $\cdot $ is kept fixed) which in physics corresponds to
deformations by open membrane operators (the closed string viewed as sitting
on the boundary of the open membrane). Finally, there is the Gerstenhaber
complex ($d$ is fixed but both $\cdot $ and $\left[ ,\right] $ are deformed)
which is conjectured to correspond to $\left( 1,1\right) $ little string
theory, i.e. strings attached to an NS5 brane background in type IIB
superstring theory (see \cite{HM} for the details).

Using the results of \cite{HM}, Lemma 2 immediately has the following
corollary, then:

\bigskip

\begin{corollary}
Let $\left( A,\cdot ,\left[ ,\right] \right) $ be the Gerstenhaber algebra
of a topological closed string BRST-complex. The deformation theory
described by the Gerstenhaber complex of \cite{HM} is completely equivalent
to the deformation theory of the odd super-trialgebra $\widehat{U}_G\left(
A\right) $ with the coproduct $\Delta $ of $\widehat{U}_G\left( A\right) $
kept fixed.
\end{corollary}

\bigskip

Observe that by construction one can - as in the case of Lie algebras -
reconstruct the Gerstenhaber- or the Poisson algebra from the given
universal envelope. Especially, we really get a 1-1 correspondence of the
deformation theories in the above corollary.

\bigskip

\begin{remark}
The corollary shows that one can generalize the deformation theory of the
Gerstenhaber algebra to the full deformation theory of the odd
super-trialgebra by including deformations of the coproduct $\Delta $.
Suppose the conjecture of \cite{HM} on a correspondence of the Gerstenhaber
complex to $\left( 1,1\right) $ little string theory on an NS5 brane in type
IIB string theory would hold true. Then, the full deformation theory of the
universal envelope $\widehat{U}_G\left( A\right) $ should correspond to a
noncommutative deformation of $\left( 1,1\right) $ little string theory
since the odd Lie-structure of the BRST-complex is replaced by a quantum
algebra in such a deformation. It remains a task for future work to check if
this produces exactly the noncommutative deformations of little string
theory which are known to appear in the presence of a stack of NS5-branes
(see \cite{GMSS} and \cite{Har}).
\end{remark}

\begin{remark}
The above corollary means that the deformation theory of Gerstenhaber
algebras, enlarged as just pointed out to the full deformation theory of $%
\widehat{U}_G\left( A\right) $, should have strong stability properties
corresponding to ultrarigidity which holds for odd super-trialgebras
completely analogous to the case of trialgebras considered in \cite{Sch}.
\end{remark}

\bigskip

We conclude this section by pointing out a possible application of the
universal envelope construction of a Gerstenhaber algebra: It suggests that
one can approach the proof of the conjecture of \cite{HM} on a relation of
the Gerstenhaber complex to $\left( 1,1\right) $ little string theory in a
representation theoretic way, at least in a special case.

Consider the NS5 brane configuration in type IIB string theory as used in 
\cite{LMN} where the $\left( 1,1\right) $ little string theory is suggested
to correspond by S-duality to instantons on the S-dual D5. It was shown in 
\cite{LMN} that these can be related by another duality transformation to
the A-model on $\Bbb{CP}^1$ with gravitational descendants taken into
account. It was further shown in \cite{EHY} that the A-model on $\Bbb{CP}^1$
can - including the case of coupling to topological gravity - be related by
mirror symmetry to a B-model, i.e. we can add another segment to the chain
of dualities starting from the $\left( 1,1\right) $ little string theory on
the NS5 in type IIB. Concretely, the B-model is given as a Landau-Ginzburg
model with potential given by an $N=2$ supersymmetric sine-Gordon model.

For simplicity, let us consider the usual (bosonic) sine-Gordon model and
dispense of the coupling to topological gravity. It has been known for a
long time that the fusion ring of the sine-Gordon model cannot - in contrast
to the case of two dimensional conformal and three dimensional topological
field theories - be generated by the representation theory of Hopf algebra
deformation of a compact group (see \cite{FK}, \cite{LR}). Indeed, it can be
shown that the hidden symmetry of the sine-Gordon model is given by a
quantum deformation of the affine Lie algebra $\widehat{sl}_2$ with the
restricted sine-Gordon model being related to the quantum deformation of the
Virasoro algebra of \cite{FR} (see \cite{FL}). For the family of restricted
sine-Gordon models it was shown (see \cite{FL}) that the conformal limit is
given by the minimal unitary series and that the deformation from the
minimal model to the corresponding restricted sine-Gordon model (leading to
the deformation of the Virasoro algebra) is generated by the $\phi ^{\left(
1,3\right) }$ field of the minimal model, i.e. it corresponds to the
deformation 
\begin{equation}
S\mapsto S+\frac \lambda {2\pi }\int d^2z\ \phi ^{\left( 1,3\right) }\left(
z,\overline{z}\right)  \label{3}
\end{equation}
of the action.

Now, by definition, a deformation as in (\ref{3}) is of WDVV-type, i.e. if
the conjecture of \cite{HM} holds true it can not be traced back along the
above chain of dualities to the same complex on the $\left( 1,1\right) $
little string theory side (because, there, the Gerstenhaber complex is
expected to appear). In contrast, the fact that the deformations arising
from turning on the background fields in $\left( 1,1\right) $ little string
theory can definitely not be described as WDVV-type deformations shows that
the three bicomplexes introduced in \cite{HM} are not kept separate from
each other under dualities. Trying to prove the conjecture of \cite{HM} by
applying the above chain of dualities leads, especially, to the following
question, then: Can we find a nontrivial odd-supertrialgebra relating to the
sine-Gordon model?

Remember that by Kazhdan-Lusztig theory the representation theory of an
affine Lie-algebra corresponds to the deformation theory of a compact
quantum group. One might suspect that the same holds true one level of
deformations higher: The representation theory of a quantum affine algebra
should correspond to the representation theory of a compact trialgebra
(trialgebraic deformation of a compact quantum group). We therefore
conjecture that it should be possible to prove the conjecture of \cite{HM}
by proving that the fusion structure of the sine-Gordon model corresponds to
the representation theory of an odd-super trialgebra and that this result
persists to hold true for the $N=2$ sine-Gordon model coupled to topological
gravity.

We expect the appearance of an odd super-trialgebra - instead of a usual
bosonic trialgebra - even for the case of the simple (bosonic) sine-Gordon
model because of the enlarged set of generators appearing for the $q$%
-deformed Virasoro algebra (see \cite{FR}).

The following lemma provides support for our above conjecture:

\bigskip

\begin{lemma}
On the cohomology of any affine Hopf algebra there is up to homotopy an
action of the Hopf algebra $\mathcal{H}_{GT}$ of \cite{Sch}.
\end{lemma}

\proof%
The analogous action of the Grothendieck-Teichm\"{u}ller group $GT$ on
Hochschild cohomology is proved in \cite{KS}. Using this result, ignoring
compatibility of the product and coproduct of a Hopf algebra, we would get
an action of $GT$ on the cohomology of the product and a coaction of the
Hopf algebra dual of $GT$ on the cohomology of the coproduct. Extending the
action of $GT$ by linearity to one of its group algebra, we arrive at an
action of the Drinfeld double $\mathcal{D}\left( GT\right) $ of $GT$ on the
complete cohomology (involving product and coproduct). One checks by
calculation that imposing the compatibility between product and coproduct
still respects the defining equations of a sub-Hopf algebra. So, we arrive
at a sub-Hopf algebra $\mathcal{H}$ as acting on the cohomology of affine
Hopf algebras. By direct calculation one checks also that 
\[
\mathcal{H\simeq H}_{GT}
\]
This concludes the proof. 
\endproof%

\bigskip

\begin{remark}
As we see from the proof it holds in greater generality than for affine Hopf
algebras, only. We have restricted to the affine setting because e.g. for
the classical quantum groups the action of $\mathcal{H}_{GT}$ trivializes in
part because it is always possible up to isomorphism to assume that either
only deformations of the product or only deformations of the coproduct occur
(see e.g. \cite{CP}).
\end{remark}

\bigskip

In section 4, we will give an additional argument in support of the
conjecture of \cite{HM} which provides another application of our universal
envelope construction for Gerstenhaber algebras.

\bigskip

\section{Deformations of Poisson algebras}

Obviously, one can introduce the three bicomplexes introduced in \cite{HM}
for a differential Gerstenhaber algebra also for a differential Poisson
algebra. Let us consider the case of a pure Poisson algebra, i.e. we have a
trivial differential which is not deformed. Then the three complexes
correspond to:

\bigskip

\begin{enumerate}
\item  Deforming $\cdot $ while $\left[ ,\right] $ is kept fixed.

\item  Deforming $\left[ ,\right] $ while $\cdot $ is kept fixed.

\item  Deforming both $\cdot $ and $\left[ ,\right] $.
\end{enumerate}

\bigskip

We will consider the third possibility in a number of examples, now. We will
discover that in all of these examples the third possibility is never
realized and even more severe restrictions occur. In this section, we will
sometimes omit to explicitly write the symbol $\cdot $ (as usual for a
product).

\bigskip

\begin{example}
Consider the polynomial algebra over the Euclidean plane, i.e. the real
associative unital algebra with generators $x,y$ and relation 
\[
xy=yx
\]
We can introduce the usual Poisson bracket on this algebra, i.e. we have 
\[
\left\{ x,y\right\} =1
\]
It is straightforward to check that we get a Poisson algebra in this way.
Assume we have a deformation of the algebra with commutator 
\begin{equation}
\left[ x,y\right] =xy-yx=\lambda \in \Bbb{R}  \label{4}
\end{equation}
and 
\begin{equation}
\left\{ x,y\right\} =1+\mu xy,\ \mu \in \Bbb{R}  \label{5}
\end{equation}
Observe that since the monomials $x^ny^m,\ n,m\in \Bbb{N}$ give a basis of
the algebra, any deformation of $\left\{ ,\right\} $ has to satisfy 
\[
\left\{ x,y\right\} =\sum_{n,m}a_{nm}x^ny^m
\]
By bilinearity of $\left\{ ,\right\} $ - up to the constant non-deformed
term - only (\ref{5}) remains, then. The requirement that the deformation
should constitute a Poisson algebra, again, leads to 
\begin{eqnarray*}
\left\{ xy,y\right\}  &=&x\left\{ y,y\right\} +\left\{ x,y\right\} y \\
&=&\left\{ x,y\right\} y
\end{eqnarray*}
and 
\begin{eqnarray*}
\left\{ xy,y\right\}  &=&\left\{ yx+\lambda ,y\right\}  \\
&=&\left\{ yx,y\right\} +\lambda \left\{ 1,y\right\}  \\
&=&\left\{ yx,y\right\}  \\
&=&y\left\{ x,y\right\} 
\end{eqnarray*}
i.e. 
\[
\left\{ x,y\right\} y=y\left\{ x,y\right\} 
\]
Using (\ref{5}), we get 
\begin{eqnarray*}
\mu xy^2 &=&\mu yxy \\
&=&\mu \left( xy-\lambda \right) y
\end{eqnarray*}
i.e. 
\[
\mu \lambda =0
\]
So, either $\lambda =0$ or $\mu =0$ and possibility 3.) in the deformation
theory of a Poissson algebra does not occur.. In consequence, for a
noncommutative space of type (\ref{4}) (i.e. $\lambda \neq 0$) only the
WDVV-like deformations (in the case of a Gerstenhaber algebra, they
correspond to the WDVV-deformations) with fixed Poisson bracket are
possible. Especially, this means that, up to multiplication by the constant $%
\lambda $, the Poisson bracket equals the commutator of the noncommutative
product (\ref{4}). This agrees with the usual interpretation of the Poisson
bracket in deformation theory where it gives the first order contribution
for the deformation of the product. So, starting from the more general
notion of a Poisson algebra, we are automatically forced to this
interpretation of the Poisson bracket. We will see that this holds true in
all the examples to follow.
\end{example}

\bigskip

\begin{example}
Consider, again, the Poisson algebra on the Euclidean plane. We use (\ref{5}%
) for the deformation of the bracket but instead of the Heisenberg type
deformation (\ref{4}) for the product, we use a Lie algebra type deformation
(as one does in physics e.g. on the fuzzy sphere): 
\begin{equation}
\left[ x,y\right] =\lambda _1x+\lambda _2y,\ \lambda _1,\lambda _2\in \Bbb{R}
\label{6}
\end{equation}
Using the derivation property for 
\[
\left\{ xy,y\right\} 
\]
and relation (\ref{6}), we now get 
\[
\left\{ x,y\right\} y=y\left\{ x,y\right\} +\lambda _1\left\{ x,y\right\} 
\]
and using (\ref{5}) 
\[
y+\mu xy^2=y+\mu yxy+\lambda _1+\lambda _1\mu xy
\]
i.e. 
\[
0=\mu \left( \lambda _1x+\lambda _2y\right) y+\lambda _1+\lambda _1\mu xy
\]
Using the basis property of the monomials, we get 
\[
\lambda _1=0
\]
and 
\[
\mu \lambda _2=0
\]
A similar calculation with $x,y$ interchanged leads to 
\[
\lambda _2=0
\]
So, there is no Lie algebra type deformation of $\cdot $ which is compatible
with a continuous deformation of $\left\{ ,\right\} $ or even with the
undeformed bracket. Again, we would be forced to choose the Poisson bracket
as the first order deformation of the deformed product to achieve
compatibility and can not consider $\left\{ ,\right\} $ as an independent
algebraic structure.
\end{example}

\bigskip

\begin{example}
Again, we take the Poisson algebra on the Euclidean plane and (\ref{5}). For
the deformation of the product we now take the $q$-deformation type 
\begin{equation}
xy=qyx,\ q\in \Bbb{R}  \label{7}
\end{equation}
Applying the derivation property and (\ref{7}) to 
\[
\left\{ xy,y\right\} 
\]
now leads to 
\[
\left\{ x,y\right\} y=qy\left\{ x,y\right\} 
\]
Inserting (\ref{5}), we get 
\begin{eqnarray*}
y+\mu xy^2 &=&qy+q\mu yxy \\
&=&qy+\mu xy^2
\end{eqnarray*}
i.e. 
\[
q=1
\]
So, we reach the same conclusion as in the second example.
\end{example}

\bigskip

One might ask if the impossibility to deform the full Poisson algebra
structure and the fact that one is restricted to considering the Poisson
bracket as the first order part of the deformation of the product is an
artefact of two dimensions or might possibly occur only for algebras on
finite dimensional manifolds since there the formality theorem of \cite{Kon}
applies. To show that the same effect occurs in the infinite dimensional
case, we consider another example (which is prototypical in physics).

\bigskip

\begin{example}
Consider the infinite dimensional algebra with generators $a_n,\ n\in \Bbb{Z}
$ and relations 
\begin{equation}
a_na_m=a_ma_n  \label{8}
\end{equation}
with the bracket 
\begin{equation}
\left\{ a_n,a_m\right\} =in\delta _{n+m,0}  \label{9}
\end{equation}
This is the well known algebra of Fourier coefficients for the solutions of
the wave equation with pointwise product (\ref{8}) and the usual Poisson
bracket (\ref{9}). Let us consider the following deformation of (\ref{8})
and (\ref{9}) to 
\begin{equation}
\left[ a_n,a_m\right] =a_na_m-a_ma_n=\Theta _{nm}\in \Bbb{C}  \label{10}
\end{equation}
and 
\begin{equation}
\left\{ a_n,a_m\right\} =in\delta _{n+m,0}+\sum_k\lambda _{nm}^ka_k
\label{11}
\end{equation}
with $\lambda _{nm}^k\in \Bbb{C}$. Using (\ref{10}) and the derivation
property for 
\[
\left\{ a_la_n,a_m\right\} 
\]
we get 
\[
\sum_k\lambda _{nm}^k\Theta _{kl}+\sum_k\lambda _{lm}^k\Theta _{nk}=0
\]
Choose $n=m$. Then 
\begin{equation}
\sum_k\lambda _{\ln }^k\Theta _{kn}=0  \label{12}
\end{equation}
since the $\lambda _{nm}^k$ are antisymmetric in the lower indices.
Obviously, if (\ref{12}) has a solution it can be chosen to be independent
of $l$. But then for all $l,n,k$ 
\[
\lambda _{\ln }^k=\lambda _{nn}^k=0
\]
and we arrive at the same conclusion as in the previous examples.
\end{example}

\bigskip

We conjecture that the phenomenon observed above should hold generally for
Poisson algebras: Poisson brackets should only be interpretable as first
order contributions to the deformation of the product and there should be no
nontrivial deformation theory of Poisson algebras in the sense of the third
complex above. Besides the above examples, we have the following two general
motivating arguments in support of this conjecture:

\bigskip

\begin{itemize}
\item  In the hierarchy of little disc operads (see \cite{HM}, \cite{KS})
associative algebras appear as 1-algebras, Gerstenhaber algebras as $2n$%
-algebras for $n\geq 1$, and Poisson algebras as $\left( 2n+1\right) $%
-algebras for $n\geq 1$. The deformation theory of associative algebras was
shown to be described by a string theory (\cite{CaFe}, \cite{Kon}) while the
deformation theory of a Gerstenhaber algebra is described by $M$-theory and
string theory with background fields turned on (\cite{HM}). The expectation
that $M$-theory is complete as a physical theory and needs no further
deformation is fully in accordance with the conjecture that the deformation
theory of Poisson algebras just reduces to the simpler deformation theory of
associative algebras.

\item  One could embed the hierarchy of $n$-algebras into a categorical
hierarchy of higher algebras (using a universal envelope construction of the
type given above), passing from algebras to bialgebras, to trialgebras, to
quadraalgebras (with two associative products and two coassociative
coproducts, all pairwise compatible), etc.. In this context, the triviality
of the deformation theory of Poisson algebras should be a consequence of
ultrarigidity (there are no nontrivial deformations of trialgebras into
quadraalgebras, see \cite{Sch}).
\end{itemize}

\bigskip

If our above conjecture holds true, there is a decisive difference between
the highly restricted deformation theory of Poisson algebras and the graded
case with the rich deformation theory of Gerstenhaber algebras. Considering
trialgebraic structures arising as universal envelopes of Poisson- and
Gerstenhaber algebras, this means the following: For trialgebras arising as
universal envelopes of Poisson algebras we should not expect a deformation
theory which goes much beyond the theory of bialgebras and Hopf algebras.
For odd super-trialgebras, on the other hand, we expect a rich deformation
theory of deep relevance for physics (connected to string- and $M$-theory).
In other words: The super-extension might be essential for trialgebras.

In \cite{CF} trialgebras have been introduced with the aim to construct four
dimensional topological invariants. Up to now, only very simple invariants
have been constructed in this way. There might be a lesson to draw from our
above conjecture for four dimensional topological quantum field theory:
Maybe one has to pass to odd super-trialgebras to get nontrivial invariants.
In this sense, supersymmetry might be a necessary ingredient for four
dimensional topological quantum field theory.

\bigskip

\section{A trialgebra in the WZW-model}

The deformation theory and moduli of $\left( 1,1\right) $ little string
theory is a hardly accessible topic. This is one reason why a direct
approach to the conjecture of \cite{HM} seems to be very difficult. On the
other hand, the behaviour of strings in the transversal geometry of an NS5
brane has a much more accessible description in terms of a rational
conformal field theory (\cite{CHS}, \cite{Rey}). One can therefore try to
find support for the conjecture of \cite{HM} by trying to find the algebraic
properties of the Gerstenhaber complex in string theory on the transversal
geometry. Using our results on the universal envelope from section 2, we can
more concretely phrase the following question: Can we find an odd
super-trialgebra in the string theory on the transversal geometry? We will
see in this section that the answer is in the affirmative.

For a flat NS5 brane, the background is completely determined by (see e.g. 
\cite{BS}) vanishing R-R fields and 
\begin{eqnarray*}
ds^2 &=&\eta _{\mu \nu }dx^\mu dx^\nu +e^{-2\phi }\left(
dr^2+r^2ds_3^2\right) \\
e^{-2\phi } &=&e^{-2\phi _0}\left( 1+\frac k{r^2}\right) \\
H &=&dB=-kd\Omega _3
\end{eqnarray*}
where $\mu ,\nu =0,1,...,5$ are directions tangent to the NS5 brane. Here, $%
ds_3$ and $d\Omega _3$ denote the line element and volume form,
respectively, on the $S^3$. So, in the transversal geometry of the flat NS5
there is always contained an $S^3$. Studying strings on the transversal
geometry, we will restrict to strings moving only on this $S^3$ which are
described by the $SU\left( 2\right) $-WZW model. Open strings and $D$-branes
in the $SU\left( 2\right) $-WZW model have been studied in detail in \cite
{ARS} and the $D$-brane world volume geometry was found there to correspond
to the $q$-deformed fuzzy sphere of \cite{GMS}.

Consider now the bicovariant differential calculus on the $q$-deformed fuzzy
sphere (see \cite{GMS}; \cite{KlSch} for a general introduction to
bicovariant differential calculi on quantum groups). We have the following
lemma, then:

\bigskip

\begin{lemma}
The bicovariant differential calculus on a quantum group defines the
structure of an odd super-trialgebra.
\end{lemma}

\proof%
Use the fact that the bicovariant differential calculus has the structure of
a super Hopf algebra with the product given by the tensor product (see \cite
{KlSch}). In addition, the fact that it is a bimodule over the quantum group
algebra (see \cite{KlSch}, again) gives an additional realization of the
quantum group product. Using bicovariance one checks the compatibility
relations of a trialgebra by direct calculation. The fact that it is an odd
super-trialgebra derives from the super-Hopf algebra structure of the
bicovariant differential calculus and the behaviour of the quantum group
product with respect to the grading. 
\endproof%

\bigskip

\begin{remark}
Since the structure a Gerstenhaber algebra is closely related to
BRST-cohomology, this result suggests viewing the bicovariant differential
calculus on the $q$-deformed fuzzy sphere as a way to deform the
BRST-complex along with the deformation of the algebra of functions (from
the smooth functions on the sphere to the algebra of functions on the $q$%
-deformed fuzzy sphere).
\end{remark}

\bigskip

In conclusion, we find an odd super-trialgebra related to the string theory
on the $S^3$ in the transversal geometry of the NS5 brane. This gives
additional support to the conjecture of \cite{HM}.

\bigskip

\section{Conclusion}

We have given a universal envelope construction for any Poisson- and
Gerstenhaber algebra. By studying several different examples we have
motivated the conjecture that the deformation theory of Poisson algebras
should basically be those of associative algebras with the Poisson bracket
giving the first order deformations, i.e. there should be no nontrivial
deformation theory of Poisson algebras as an abstract algebraic structure.
This is in accordance with general properties of the universal envelope
(ultrarigidity). In contrast, in the case of Gerstenhaber algebras one
expects from results in string- and $M$-theory a rich deformation theory. We
have applied our construction of the universal envelope in this case to
motivate a conjecture on the possibility of a representation theoretic proof
of a conjecture of \cite{HM} on $\left( 1,1\right) $ little string theory.
Besides this, we have used the construction to get support for the
conjecture of \cite{HM} from considering the string theory on the $S^3$ in
the transversal geometry of the NS5 brane which is described by the $%
SU\left( 2\right) $-WZW model. These different instances where odd
super-trialgebras - which give the algebraic structure of the universal
envelope of a Gerstenhaber algebra - appear suggest that these might be
algebraic structures which appear in a quite universal way in string- and $M$%
-theory.

Beyond this, Lemma 4 has the following implication: The finite deformation
theory of an affine Hopf algebra is described by two Maurer-Cartan equations
(for the deformation theory of the product and of the coproduct) and one
constraint (resulting from the compatibility requirement for the
deformations of the product and the coproduct). One could consider this
system of equations as the equations of motion of a classical field theory,
generalizing the approach of \cite{BCOV} (where such an approach is followed
for holomorphic deformations). As a consequence of Lemma 4, the action
functional of this classical field theory would have to be invariant under
the action of the Hopf algebra $\mathcal{H}_{GT}$. We plan to investigate
the properties of this field theory in more detail in future work.

\bigskip

\textbf{Acknowledgments:} We would like to thank A. Klemm, M. Kreuzer, I.
Runkel, E. Scheidegger, and V. Schomerus for discussions on the material of
this paper. In addition, KGS\ would like to thank the Erwin Schr\"{o}dinger
Institute for Mathematical Physics, Vienna, for support via a Junior
Research Fellowship during the course of this work.

\end{document}